# Detection of the primary scintillation light from dense Ar, Kr and Xe with novel photosensitive gaseous detectors

L. Periale[1,2], V. Peskov[3], P. Carlson[3], T. Francke[3], P. Pavlopoulos[2], P. Picchi [1,2], F. Pietropaolo [4,2]

[1] Torino University, Torino, Italy
[2] CERN, Geneva, Switzerland
[3] Royal Institute of Technology, Stockholm, Sweden
[4] INFN Padova, Italy

The detection of primary scintillation light in combination with the charge or secondary scintillation signals is an efficient technique to determine the events "t=0" as well as particle / photon separation in large mass TPC detectors filled with noble gases and/or condensed noble gases.

The aim of this work is to demonstrate that costly photo-multipliers could be replaced by cheap novel photosensitive gaseous detectors: wire counters, GEM's or glass capillary tubes coupled with CsI photocathodes.

We have performed systematic measurements with Ar, Kr and Xe gas at pressures in the range of 1-50 atm as well as some preliminary measurements with liquid Xe and liquid Ar. With the gaseous detectors we succeeded in detecting scintillation light produced by 22 keV X-rays with an efficiency of close to 100%. We also detected the scintillation light produced by β's (5 keV deposit energy) with an efficiency close to 25%.

Successful detection of scintillation from 22 keV gamma's open new experimental possibilities not only for nTOF and ICARUS experiments, but also in others, like WIMP's search through nuclear recoil emission.

## I. Introduction

Liquid Ar/Xe (LAr/Xe) TPC's are powerful large-mass detectors which can be used in many experiments. A relevant example is the ICARUS experiment at Gran Sasso. The same technique could be used for the nTOF experiment to perform n/gamma separation or in experiments oriented on WIMP search.

One of the most powerful concepts of the LAr/Xe TPC is the detection and processing of two signals: the primary scintillation produced by charged particles or radiation and the charge signal produced by drifting primary electrons on a system of electrodes (usually wires) immersed in the liquid. Charge multiplication around the electrodes was also investigated (see for example [1] and references there in). In some cases, instead of charge signals or together with them, secondary scintillation light can be used. The secondary scintillation light can be generated in the region of strong electric field inside the liquid (or outside in the case of the double-phase detectors). An advantage of this approach is that it gives a powerful method for particle separation, because the ratio of these two signals depends on energy and other characteristics of the particles (or photons) [2]. Earlier, expensive photo-multipliers (PMTs) with window transparent to VUV were mostly used for the detection of the primary scintillation light (see [1-3] and references there in).

The aim of the present work is to implement into TPCs a novel photosensitive gaseous detector: a wire counter or GEM/capillary-type plate combined with CsI photocathodes. We name of this device is gaseous photo-multiplier or GPM.

Such detectors have the same quantum efficiency (QE) as vacuum PMTs, but in contrast to PMTs they are cheap, compact, can be manufactured with large sensitive area, are position sensitive and insensitive to magnetic fields. GPMs with CsI and other solid photocathodes show advantages with respect to ordinary photosensitive gaseous detectors due to their good time resolution and the possibility to work at low temperatures (see [4,5] and references therein)

## II. Experimental set up

Our experimental set up is presented schematically in Fig 1. Basically it contains a scintillation chamber, separated by windows from the GPM and the PMT. In some tests no separation window was used between the scintillation chamber and GPM. Inside the scintillation chamber a source was installed: $^{241}$Am, Cd or $^{60}$Sr. In some experiments we also used $^{57}$Co placed outside the test chamber. The scintillation chamber could be pumped and/or filled with noble gases: Ar, Kr or Xe at pressure p=1-50 atm or in liquid phase. The following GPMs were used: a single-wire counter with a CsI cathode, double GEM operating in tandem, and capillary tubes operating in tandem, and a home made GEM/capillary–type detector. The latter, for reason which will be presented below, was called "optimized GMP". The single-wire counter had the diameter of the cathode cylinder of 3.8 cm, and the diameter of the gold coated tungsten anode wire of 80 μm. (a disc ~2 cm diameter). To achieve the highest possible QE its design has a special feature: the disc (diameter of 22mm) covered with a CsI layer was placed on a distance of 8mm from the anode wire. Two sizes of GEM's were used: 3x3 cm2 and 10x10 cm2. In all measurements they were use in tandem. Two types of configurations were tested; either they were combined with CsI photo-cathodes placed few mm apart from the first GEM, or the upper GEM was covered with CsI photosensitive layer. Capillary plates had a diameter of 20 mm and the size of holes of 100 μm. Similarly to GEM they operated in tandem and were combined with CsI photo-cathodes. The picture of the "optimized GPM" is presented in Fig 2. It was made from G10 plate, 2 mm thick with drilled holes, 1mm diameter each. The detector was operated in a single step configuration only. Most CsI photo-cathodes used were prepared at CERN [6] however home–made sprayed photo-cathodes were also used. To avoid photo-chemical reaction of CsI with the cupper the GEM and "optimized GPM" electrodes were covered in advance with Ni/Au layers. All detectors operated in P10 or Ar+5%CH4 gas. A charge-sensitive amplifier Ortec 142 AH was connected to the anode wire via a de-coupling capacitor. When necessary, the signals from the charge-sensitive preamplifier were additionally amplified by a research amplifier Ortec 450. The GPM also had a small Be window allowing to use an X-ray source for amplitude calibration.

The PMT we used was EMI-9426 with a MgF2 window. Signals from the GPM and the PMT were recorded on a Le Croy digital oscilloscope.

The distances between the Am-source, the CsI photo-cathode and the PMT were 5 cm and 6.5 cm, respectively. The typical solid angle for detection in measurements with single wire counter was $\Omega \sim 10^{-2}$. In the case of other types of GPM, due to their lower efficiency we worked with increased solid angles (up to $\Omega \sim 10^{-1}$).

## III. Results
### 1. Detectors with windows, operating in quenched gases
#### a) Results with single wire counter
#### a-1) Results at 1 atm

The first tests were done with a strong source of primary scintillation, 241 Am, to measure the GPM's quantum efficiency. In order to estimate the number of photo-electrons emitted from the CsI photo-cathode by individual scintillation bursts produced by alpha particles, we used an external $^{55}$Fe source [7,8]. Each 5.6 keV monochromatic X-ray photon produces ~220 primary electrons inside the GPM volume. By comparing the $^{55}$Fe pulses with the scintillation pulses one concluded that the scintillation light from alpha particles, produced around 200, 100 and 20 primary electrons from the CsI photo-cathode in Xe, Kr and Ar respectively. The quantum efficiency of the CsI cathode deduced in these measurements was QE~ 20% at 175 nm.

In the next set of measurements we tried to detect scintillation light produced by 22 keV X-rays from Cd (see Fig 3). The probability of detection was then estimated from the ratio of the measured number of counts, n, from those which one can expect from the known source intensity, N, [7,8]:

$$\xi = n/N \quad (1)$$

Detection probability $\xi$ close to 100% was reached for 22 keV X-rays in Xe. From similar measurements we found $\xi \sim 50\%$ in Kr and $\xi \sim 10\%$ in Ar.

As a final step we tried to detect scintillation light produced by βs from $^{90}$Sr (MIP). To minimize the scattering we strongly collimated the Cd source and oriented it parallel to the window at a distance of 5 cm. The typical signals and pulse height distribution are shown in Fig 4. The efficiency estimated with the method described above was ~ 25%.

### a-2) Results obtained with dense Xe

Fig 5 shows the intensity of the primary scintillation light as function of Xe density. One sees that in liquid Xe, compared to gaseous Xe at 1 atm the signal drops only by a factor of two.

### b) Results with GEM and capillaries

Intense efforts are underway to use micro-pattern detectors in combination with solid photo-cathodes (see for example [9,10] and references therein).

As a comparison we also tested the operation of double GEM and double capillary plates, combined with photo-cathodes placed few mm apart form these detectors, as well as the configuration where their upper cathodes were covered with CsI [8]. Results obtained with micro-pattern detectors can be summaries as follows: we succeeded in detecting primary scintillation light produced by $\alpha$'s; compared with single wire counters, the total gain was lower by an order of magnitude (see Fig 6) and the efficiency $\eta$ ($\eta$ defined as efficiency per unit detector area and solid angle) was lower by a factor of 3-4 [11].

These tests show that, in the case of window approach, micro-pattern detectors do not offer any advantage compared to single-wire (only complications because micro-pattern detectors should be operate in multi-step configuration) especially for the applications in TPCs where there is no need for extremely high position resolution.

## 2 Windowless detectors operating in pure noble gases

As mentioned in the introduction, to increase the sensitivity of the TPC one has to work in charge multiplication mode. Charge multiplication in pure noble liquids or gases is a difficult task. This is why the use of secondary scintillation light was suggested. The secondary scintillation light can be produced near the electrodes in liquid or gas phase (the so called double-phase TPC). In the latter case, for detection of primary or secondary scintillation light it would be convenient to use windowless detectors. Micro-pattern detectors may have advantages because they can operate in pure noble gases. This is because possible photon feedback is geometrically blocked. To verify this possibility we performed several tests with GEMs, capillary tubes and optimized GPM.

### a) Windowless GEM and capillary tubes

We made short tests of operation in pure noble gases of GEMs [11] and capillary tubes with CsI photo-cathodes placed 3 mm apart as well as with capillary tubes whose upper cathodes was covered with CsI. Results obtained can be summaries as follows.

1) All detectors worked in pure noble gases, however two or more steps of multiplication were required to get gains $>10^3$ (see [11]). This makes the detection system too complicated.

2) The dynamic range of the multi-step configuration was narrow, due to limits on the maximum allowed total charge in the avalanche before break-down appears (see[12]). This restricts the dynamic range and reduces the sensitivity of the TPC.

3) The detection efficiency of the windowless GPMs η, was measured with primary scintillation produced with alphas. In Ar the detection efficiency was 2-2.5 less than the detection efficiency of a single wire counter. However in Xe and Kr it was lower by a factor of 8-15.

Therefore, this type of windowless detector can compete with a single wire counter with a window only if the window is made of CaF. In the case of MgF2 window (which is more transparent to Ar scintillation light), the single wire counter will be superior in gain, efficiency and maximum achievable gain.

### b) Tests of optimized GPM

From [12] it follows that several thin micro-pattern detector operating in tandem are equivalent in maximum achievable gain to one detector with a thick avalanche gap. This is why we developed and extensively tested an optimized design of GPM with a combination of capillary tubes and GEM (see Fig 2).

This detector allowed us to get rather good results. Without CsI coating it can operate at gains close to $10^4$, but with CsI the maximum achievable gain lowers to $10^3$ due to ion feedback.

Two possible functions of optimized GPMs were studied. The first was to use a windowless GPM for detection of primary and secondary scintillation light. Fig 7 shows pulses from a GPM and a PMT produced by the scintillation light produced by 122 keV X-rays. The detection efficiency η was 15 times lower compared to a single wire counter.

However, in practice, since the detector has a very simple and rigid design, one can compensate its low value of η by using a large surface detector. In this case the total efficiency is

$$\Sigma = \eta\, S\Omega \quad (2)$$

where S is the detector sensitive surface and $\Omega$ is the detection solid angle that can be made sufficiently large even larger than a single wire counter with window can offer.

The second possible application of the optimized GPM was to use it as primary scintillation light amplifier (see[13]). The main advantage of this approach is that in contrast to charge multiplication it has a linear response. The disadvantage is that one has to use another detector, for example with window, to detect the amplified light.

## IV. Discussion-

### 1. Detectors with window

Results presented in this work demonstrated that the scintillation light from noble gases could be successfully detected by GPMs. One of the best options among them is a single–wire counter. These detectors are very simple and cheap. They are sensitive to single photo-electrons and at the same time practically have no "dark current" or "spurious" pulses (typical of PMTs), and therefore it is beneficial in the search for very weak signals.

As a consequence, we succeeded to detect scintillation light produced by αs, 60 keV and 22 keV X-rays, with detection probability, ξ, close to 100%. With the probability of ~25% we detected the scintillation light produced by βs with 5 keV deposit energy.

Since the detected signals were at the level of single electrons, no energy resolution was obtained for 22 keV X-rays. Single electron signals is a result of the small solid angle for the scintillation light detection ($\Omega \sim 10^{-2}$). With a larger solid angle or by using reflective coating in the scintillation chamber, better results are expected.

The most important result obtained in these measurements was that the primary scintillation light does not drop significantly with pressure. This opens the possibility to use primary scintillation in high pressure detectors (similar to what was used in the case of liquid noble gases).

Once more our experiments demonstrated the superiority of GPMs with respect to PMTs: the GPM has the same quantum efficiency of the PMT, but no noise pulses or after-pulses. Therefore the use of GPMs for "high rate" application, for example nTOF, is crucial.

**2. Windowless detectors**

Results can be explained from formula (2) in which

$$\eta = QckT \quad (3)$$

where c is an extraction efficiency compared to vacuum (it depends on the gas, it is higher in Xe, lower in Ar), k is a collection efficiency from the photo-cathode surface, T is the window transparency (in the case of GPMs with window).

In the case of single-wire counters, c=0.5-1 (depending on applied voltage) and k=1.

In the case of micro-pattern detectors with CsI deposited directly on their cathode (GEM, capillary tubes, optimized GPM), k<1 and c=0.5-1 (if they operate in quenched gases) and c<<1 if they operate in noble gases.

Therefore, in the case of Ar the total efficiency $\Sigma$ of the single wire counter with window (low T) can be close to the total efficiency of the windowless GPMs (high T, but low k and c). In the case of Kr and Xe the efficiency $\eta=Qkc$ of the "optimized GPM" was lower than single wire detector with window. However, it could be compensated by using large sensitive surface. Finally we conclude that "optimized GPMs" can compete with single wire counters in some cases.

**VI. Conclusion**

For the first time the scintillation light from Ar, Kr and Xe was detected with a GPM. This may open a possibility to use simple and cheap readout of high pressure or noble liquid scintillation detectors. Advantages of GPMs compared to PMTs are: low cost, large area, insensitivity to magnetic fields, as well as very simple and compact designs. This technique can be applied not only in some important physics experiments, like WIMPS search, but also in medicine and industry to improve the time and position resolution of various detectors.

**Acknowledgments.**

We thank I. Crotti for technical assistance and the C. Williams , Sauli and  Braem group for friendly help. We would like also to thank L. Ropelewski for often discussions and help through this work

**Figure captions**

Fig. 1  Schematic drawing of the experimental set up.

Fig. 2 Photo of optimized GPM.

Fig. 3 Oscillograms of the signals from the GPM (the upper beam) and the PMT (the meadle beam) in the case of the detection a primary scintillation light produced by 22 kev X-rays in Kr at a pressure of 1 atm. A pulse-height spectrum is also shown (the lowest beam)

Fig. 4 Oscillograms of the signals from the GPM and the PMT in the case of the detection of the primary scintillation light produced by beta particle with 5 kev deposit energy. Xe, 1 atm.

Fig. 5 Dependence of the intensity of the primary scintillation light (produced by 30 keV photons) with Xe density. Note that in the case of LXe a 60keV souse was used

Fig. 6 Gain of double-step GEM operating in P10 gas at 1atm. Diamond- without CsI coating, squares- with CsI coating

Fig. 7 Signals from the optimized GMP and PM produced by scintillation light from Xe at p=2.5 atm by 122keV gamma-rays.

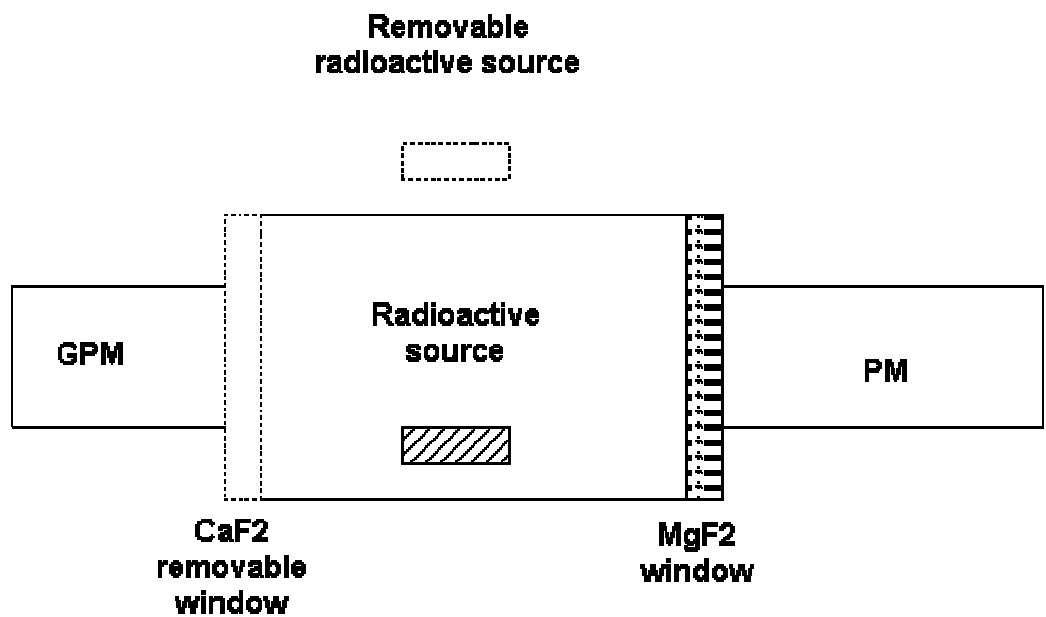

Figure 1

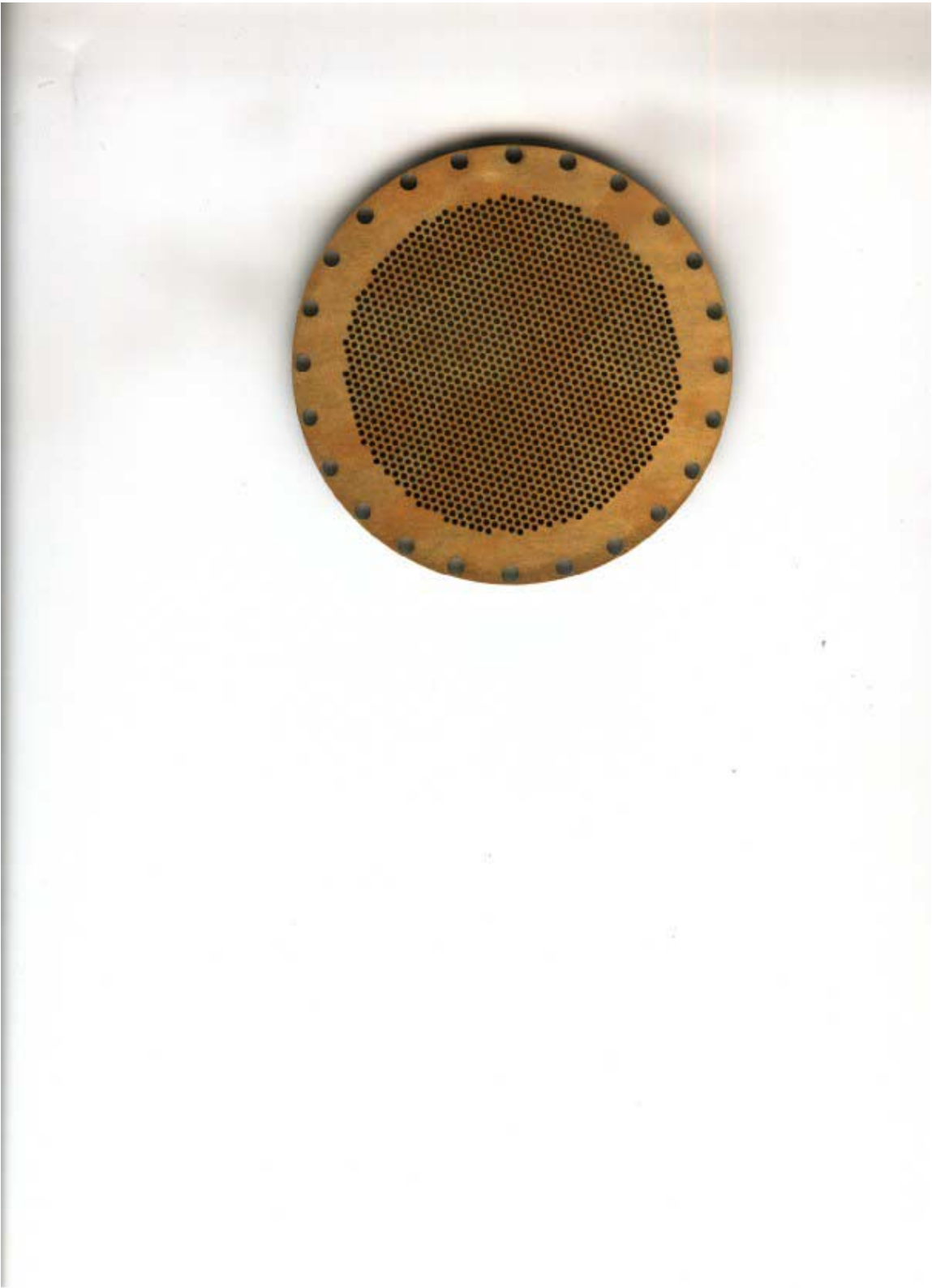

Figure 2

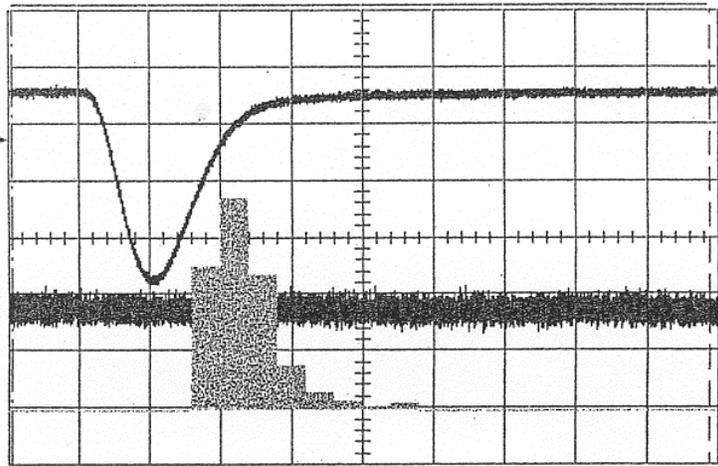

Figure 3

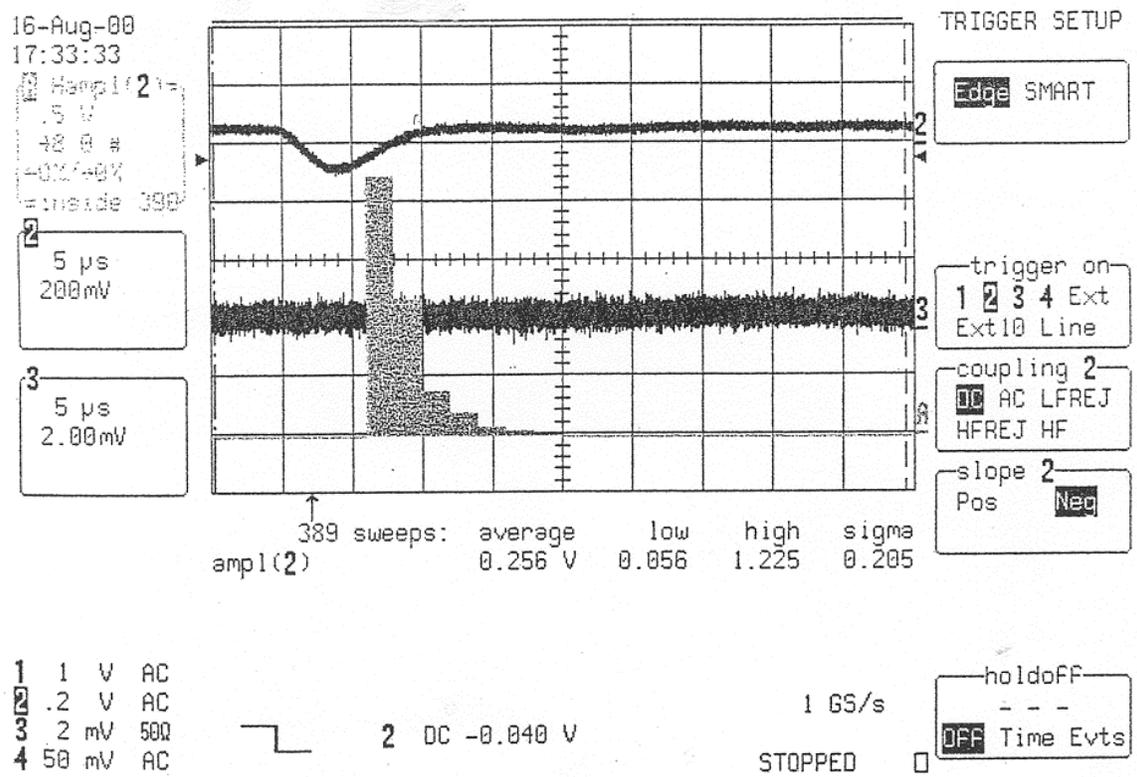

Figure 4

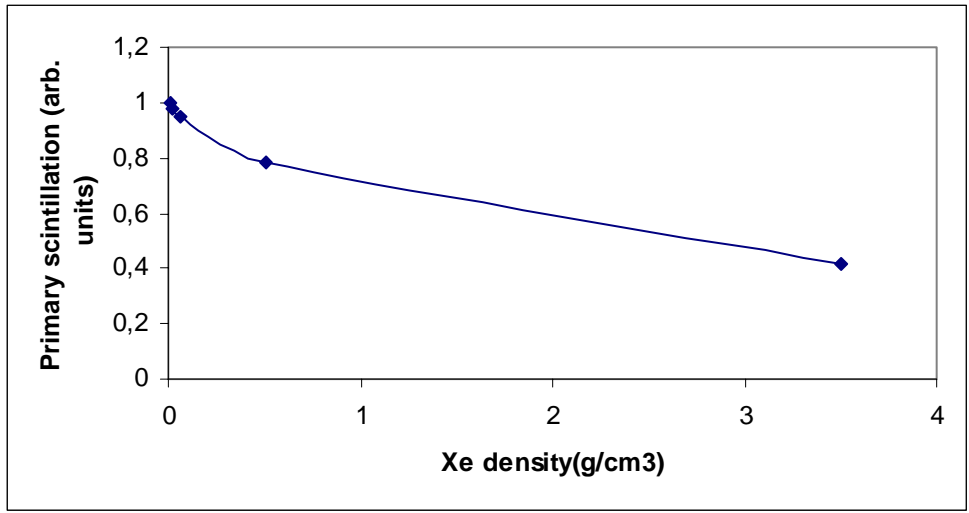

Figure 5

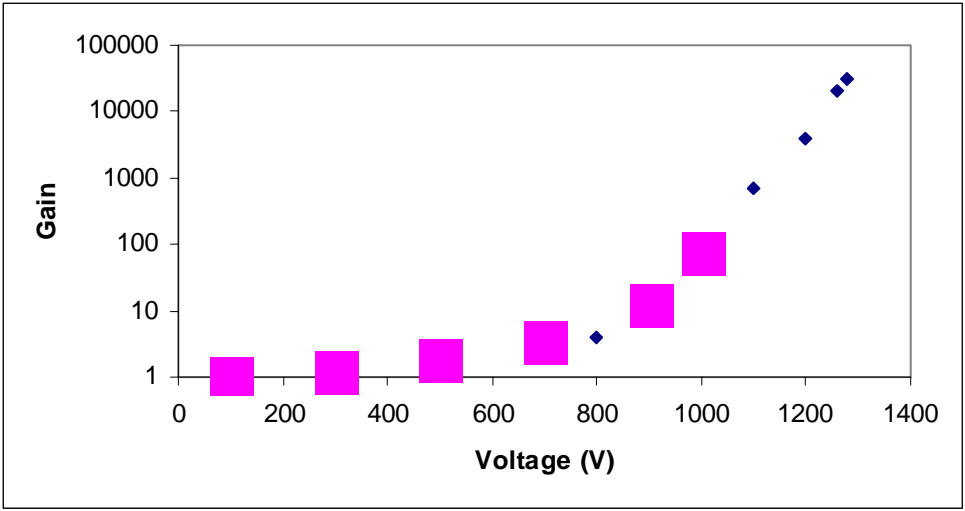

Figure 6

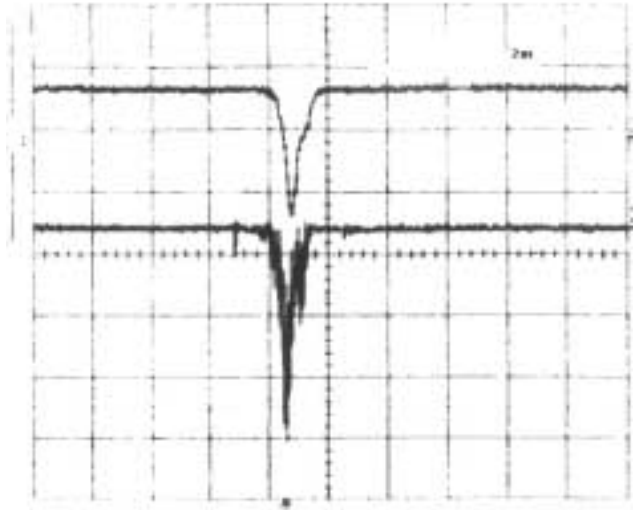

Figure 7